\documentclass[11pt]{article}
\usepackage[utf8]{inputenc}

\usepackage{setspace}
\usepackage[margin=1in]{geometry}
\usepackage{bm}
\usepackage{graphicx}

\usepackage{booktabs}
\usepackage{siunitx}
\newcolumntype{d}{S[
    input-open-uncertainty=,
    input-close-uncertainty=,
    parse-numbers = false,
    table-align-text-pre=false,
    table-align-text-post=false
 ]}

\providecommand{\keywords}[1]
{
  \small	
  \textbf{\textit{Keywords: }} #1
}

\usepackage{natbib}
\setcitestyle{authoryear,open={(},close={)}}

\usepackage{hyperref}

\title{COVID-19 Demand Shocks Revisited: Did Advertising Technology Help Mitigate Adverse Consequences for Small and Midsize Businesses? \vspace{1cm}}
\author{Shun-Yang Lee (Northeastern University), Julian Runge (Northeastern University),\\ Daniel Yoo (Meta Platforms), Yakov Bart (Northeastern University),\\ Anett Gyurak (Meta Platforms), J.W. Schneider (Meta Platforms)\footnote{The first three authors contributed equally and are listed in alphabetical order. Address correspondence to j.runge@northeastern.edu or sh.lee@northeastern.edu.}}

\date{January 17, 2024}

\begin{document}

\begin{titlepage}

\maketitle
\thispagestyle{empty}

\vspace{0.4cm}

 \begin{abstract}
    	\noindent Research has investigated the impact of the COVID-19 pandemic on business performance and survival, indicating particularly adverse effects for small and midsize businesses (SMBs). Yet only limited work has examined whether and how online advertising technology may have helped shape these outcomes, particularly for SMBs. The aim of this study is to address this gap. By constructing and analyzing a novel data set of more than 60,000 businesses in 49 countries, we examine the impact of government lockdowns on business survival. Using discrete-time survival models with instrumental variables and staggered difference-in-differences estimators, we find that government lockdowns increased the likelihood of SMB closure around the world but that use of online advertising technology attenuates this adverse effect. The findings show heterogeneity in country, industry, and business size, which we discuss and is consistent with theoretical expectations.
  \end{abstract}

\vspace{1cm}


\vfill

  \noindent \footnotesize{}

\end{titlepage}

\doublespacing

\section{Introduction}

The COVID-19 pandemic posed a serious threat to the survival of businesses around the world. As the pandemic unfolded, businesses were faced with the dire consequences of falling demand, declining revenues, and liquidity constraints. Indeed, as the pandemic intensified, businesses closed at an alarming rate \citep{apedo2020unmasking}.\footnote{Apedo-Amah et al. (2020) find that the likelihood of business closures within four weeks of the peak of the COVID-19 pandemic around the world was more than 70\%.} While the pandemic has increased researchers' interest in understanding the extent and moderators of these closures \citep{bartik2020impact, bartlett2020small,gourinchas2020covid, wang2020bankruptcy}, few insights exist into how online advertising technology may have shaped the performance and survival of businesses around the world.

One reason for this lack of insight on the interaction of COVID-19 demand shocks, business survival, and advertising technology use is the paucity of data that observe both the advertising and survival of businesses during the pandemic. Although there has been a critical need to understand how businesses have been impacted by the pandemic, the sudden and rapid onset of COVID-19 has made it difficult to obtain timely and detailed data, especially outside a small group of regions and countries. Moreover, recent efforts have focused on collecting data on a basic set of financial and employment indicators \citep{apedo2020unmasking, bartik2020measuring,chetty2020real}. As such, the availability of advertising data has been limited by comparison, and, to the authors' best knowledge, no studies have closely examined the role of advertising technology use in business survival during the pandemic.

In this research, we construct and analyze a novel data set to address this gap. Using data of more than 60,000 businesses collected in 49 countries on the Facebook advertising platform from May to October 2020, we examine their use of advertising technology and survival during the pandemic. Building on recent studies, we first investigate how government interventions in response to the pandemic have affected the survival of businesses around the world. We then turn to businesses' use of advertising technology to explore whether different aspects of personalized online advertising have moderated the impact of COVID-19 demand shocks from government interventions. We extend this analysis by exploring heterogeneous effects that might exist across several characteristics, including region, industry, and business size. Finally, we run a series of additional analyses to test the robustness of our results. Our interest is in the effects among small and midsize businesses (SMBs), which were most adversely affected by the pandemic \citep{alekseev2023effects}.

Using a set of discrete-time survival models with instrumental variables, we find that government lockdowns during the pandemic led to an increase in business closures around the world, but that online advertising moderated the impact of these lockdowns. We show that an average increase in lockdowns in the first month of the pandemic was associated with an average decrease in survival probability of 8.3 percentage points across all countries in our sample. However, we also find that a one standard deviation increase in online advertising expenditures was associated with an average increase in survival probability of 2.8 percentage points across all countries.  Additional analyses using a staggered difference-in-differences estimator corroborate that the use of advertising technology can attenuate negative effects of government lockdowns on business survival. In addition to these aggregated findings, we uncover substantial heterogeneity across regions, countries, industries, and business sizes. We find that SMBs in the services, transportation, and hospitality industries were the most severely affected by government lockdowns, but they also attained some of the greatest benefits from online advertising. This was also the case for small businesses, including microbusinesses with fewer than 10 employees. Moreover, SMBs in the agricultural, manufacturing, and construction industries and larger SMBs with 500 or more employees were the least affected by lockdowns but achieved the smallest gains from online advertising. 

Our findings provide several important insights. First, in terms of business outcomes during the COVID-19 pandemic, we shed light on the role of online advertising technology that can help companies reach tailored customer segments with personalized messages quickly and at scale. While prior studies have examined how business outcomes, such as revenues \citep{chetty2020real,bloom2021impact}, layoffs \citep{bartik2020impact, palacios2021early}, and business closures \citep{apedo2020unmasking, crane2022business}, were affected during the pandemic, few studies have examined the extent to which advertising shaped these outcomes. Our findings suggest that advertising played an important role in supporting SMBs in navigating the hardships induced by the pandemic and related government measures. Second, we focus specifically on the role of personalized online advertising technology and contribute to the growing body of literature on its relevance for and impact on businesses. Research has investigated the theoretical foundations and consequences of this form of advertising \citep{lambrecht2013does, bleier2015personalized,boerman2017online}, but this work provides new evidence on its moderating role in the impact of severe demand shocks on business closures. Third, we build on recent studies on the effects of government responses to the pandemic and uncover their impact on SMBs from a cross-country perspective. Prior studies on the impact of stay-at-home orders have largely focused on a single country or region \citep{bartlett2020small,aum2021covid,goolsbee2021fear,alexander2020stay}. By contrast, we provide more general evidence across multiple regions and countries. 

We proceed as follows. Section 2 reviews the related literature on the impact of COVID-19 and identifies several gaps in research to date. Section 3 discusses the research on the effects of advertising and business closures and proceeds to outline our theory and hypotheses. Sections 4 and 5 discuss our data and methods, and Section 6 presents our main results. Section 7 presents the results of our additional analyses, and Section 8 summarizes our contributions before concluding. 

\section{Research on the Impact of the COVID-19 Pandemic}

The COVID-19 pandemic has led to a growing body of research examining how businesses have been affected around the world. Researchers have been deeply interested in understanding both the consequences of the pandemic for businesses and the drivers and moderators of these consequences, given the devastating toll of the pandemic on human life and the fundamental changes it has brought for society and the economy.

Researchers who collected data during the early stages of the pandemic were the first to signal the grave consequences of the pandemic for businesses. In a survey of US SMBs between late March and early April 2020, \citet{bartik2020impact} found that mass layoffs and business closures occurred just a few weeks into the pandemic, with more than 40\% of businesses in their sample closing. In another survey conducted from late March to April, \citet{humphries2020evolving} found that more than half of US SMBs had already laid off a substantial portion of their employees.

As the pandemic intensified, researchers accumulated further evidence of the serious consequences for businesses. \citet{bloom2021impact} found that US SMBs experienced an average sales decline of 29\% from March to August 2020, with small and offline businesses hit the hardest. \citet{apedo2020unmasking} echoed this finding across 51 countries, showing an average sales reduction among businesses of 49\% between April and August 2020, though considerable heterogeneity exists across countries and industries. At the same time, \citet{bartik2020measuring} found that hiring among large and small businesses in the United States began to recover from mid-April to mid-June but declined thereafter as the virus surged. Outside the United States, \citet{apedo2020unmasking} found that 57\% of businesses reported employment reductions across 51 countries between April and July 2020. Moreover, \citet{crane2022business} showed that business closures in the United States were elevated from March to September, especially among small businesses and those sensitive to social distancing rules. More recently, \citet{alekseev2023effects} used surveys fielded on Facebook to show that the pandemic had an adverse effect on small businesses, with the effect being heterogeneous across business characteristics.
	
Researchers have therefore provided valuable findings on the negative consequences of the pandemic for businesses. In addition, they have worked to better understand the reasons behind these outcomes. Notably, the role of government lockdowns and their impact on the performance and survival of businesses have been a major topic of interest. Although researchers have tried to explain why stay-at-home orders might have adversely affected businesses during the pandemic, disentangling different components of the effect is not straightforward. On the one hand, \citet{alexander2020stay} and \citet{gourinchas2020covid} find that stay-at-home orders caused considerable reductions in consumer spending and led to increased failures among small businesses during the pandemic, respectively. \citet{apedo2020unmasking} also find that business closures around the world were at their highest in the weeks following the most stringent periods of lockdown. On the other hand, \citet{chetty2020real} find that stay-at-home orders had little or no impact on consumer spending, business closures, or employment. Instead, they find that the primary barrier to economic activity was voluntary self-limiting due to the threat of COVID-19 itself. Moreover, \citet{goolsbee2021fear} find that stay-at-home orders accounted for part of the decline in economic activity but that reductions in mobility were more closely tied to the fear of infection.

Research on business outcomes during the COVID-19 pandemic has therefore rapidly grown since its onset. While studies have worked to uncover how the pandemic affected businesses and through which mechanisms, several important questions remain unanswered. First, how did advertising influence businesses during the pandemic? Prior research suggests that advertising should matter for both the performance and survival of businesses during the pandemic \citep{sethuraman2011well,jindal2015impacts}, and \citet{alekseev2023effects} suggest that advertising may have mitigated the adverse effect of the pandemic on businesses. Yet, few studies have formally investigated these relationships. Second, to what extent can the research thus far be generalized to other places and contexts? While research has tried to understand the consequences of the pandemic in specific countries, whether these findings are consistent across countries, regions, and industries is relatively unknown. We therefore aim to address these questions and offer new insights on the role of advertising for businesses during the pandemic across a global sample.

\section{Advertising and Business Survival during the COVID-19 Pandemic}

Recent studies show that the COVID-19 pandemic led to substantial reductions in business sales and widespread layoffs and business closures. However, this does not tell the whole story. Advertising can be expected to have played a critical role for both the performance and survival of businesses during the pandemic for both theoretical and empirical reasons. Notably, marketing theory posits that advertising activities should have a positive effect on expected future cash flows by supporting the development of market-based intangible assets \citep{srivastava1998market, mcalister2016advertising}. Specifically, advertising helps raise consumer awareness of a business, its products, and its distinguishing features, which supports the creation of market-based intangible assets, such as brand equity and customer relationships \citep{keller1993conceptualizing, srivastava1998market}. These market-based assets built through advertising then help reduce the volatility of a business’s expected future cash flows \citep{joshi2010direct, sethuraman2011well}. 

Corroborating these predictions, many empirical studies have found that past advertising enhances the financial performance of businesses. A collection of meta-studies finds that advertising expenditures have a positive impact on sales \citep{sethuraman2011well}, profitability \citep{rubera2012firm}, and firm value \citep{edeling2016marketing}, though with noteworthy differences across industries. Studies have also found that advertising has an important influence on business survival, with past advertising both reducing the risk of bankruptcy \citep{jindal2015impacts} and increasing the likelihood of surviving the bankruptcy process \citep{jindal2020impact}. These benefits tend to persist during periods of economic downturn, with advertising lessening reductions in profitability during economic contractions for volatile industries \citep{steenkamp2011impact} and lowering bankruptcy risk in times of market turbulence \citep{jindal2015impacts}. Recent studies also highlight that marketing investments during and after the onset of an economic downturn can be beneficial to businesses \citep{angulo2022marketingfundingrecession,srinivasan2011advertisingrecession}, but firms often cut back on these investments instead \citep{oezturan2014downturnturkey,rollins2014downturnmarketing}.

These studies serve as a foundation for understanding the expected effect of advertising on businesses during the pandemic. Businesses that engaged in advertising should have been more likely to accumulate market-based assets that supported their financial performance during the pandemic, increasing their likelihood of survival. Therefore, advertising should have supported the resilience of these businesses. However, several outstanding features of the pandemic raise additional questions. Notably, the pandemic led to a major shift in the way businesses around the world advertise. Internet usage soared during the pandemic \citep{feldmann2020lockdown, koeze2020virus}, and e-commerce closely followed \citep{alfonsocommerce}. In response, businesses around the world have increasingly moved to online forms of advertising, which has accelerated a trend that was already in motion \citep{boerman2017online,gordon2019advertisingexperiments,li2020covid}. This raises the question whether online advertising should have affected the survival of businesses in the manner discussed previously.

Research on the impact of online advertising for businesses has expanded in recent years. This body of research has focused especially on the question of how consumers respond to online advertisements that incorporate different levels of personalization, including the effect of online advertising that targets on the basis of an individual’s online behavior, as well as the subsequent effect of these advertisements on business outcomes \citep{boerman2017online,gordon2019advertisingexperiments,gordon2023nonexperimental}. Foundational research on advertising suggests that higher levels of personal relevance increase the effectiveness of advertising \citep{greenwald1984audience, celsi1988role}. However, researchers also note that consumers may not respond favorably to personalized ad content \citep{smit2014understanding,o2015people} and that the effectiveness of these advertisements may depend heavily on their timing and placement \citep{bleier2015personalized}.

Empirical research thus far suggests that personalized online advertising leads to benefits that are largely consistent with those discussed previously. Studies have found that personalized online advertising can enhance market-based intangible assets, including brand equity, by enhancing brand identification \citep{tran2020personalized,gordon2023nonexperimental}. Personalized online advertising also has a positive impact on brand-specific page views \citep{rutz2012does} and both active and passive searches on a business’s web page \citep{ghose2016toward}. Moreover, personalized online advertisements have a positive impact on both revenues and repeat purchase probabilities for businesses \citep{manchanda2006effect,lewis2014online}. In addition, research has shown that the impact of personalized online advertising can be conditional on both the features of the consumer and those of the advertisement. For example, advertisements with higher levels of personalization are more effective at eliciting a response from consumers, but they also lose their potency over time \citep{bleier2015personalized}. Repeatedly targeted advertisements tend to be more effective at increasing purchases when consumers have narrowly construed preferences \citep{lambrecht2013does}, and targeted advertisements lose their effectiveness when paired with obtrusive features such as pop-ups and videos \citep{goldfarb2011online}. Nonetheless, existing research suggests that personalized online advertising helps enhance the performance of businesses overall.

These insights provide a framework for understanding how online advertising should have affected the survival of businesses during the COVID-19 pandemic. When combined with the findings of research on effects of the pandemic, they lead to three main hypotheses. First, we hypothesize that government lockdowns during the COVID-19 pandemic had a deleterious effect on businesses around the world, decreasing their chances of survival. Existing studies have explored how government lockdowns affected business closures, but the results are mixed, and we revisit this hypothesis beyond the context of the United States to understand how these lockdowns affected business closures around the world. Second, we hypothesize that online advertising moderated the adverse effects of government lockdowns. Specifically, online advertising before the pandemic should have better enabled businesses to foster market-based assets that enhanced and reduced the volatility of their cash flows during the pandemic, which in turn should have increased their likelihood of survival. Third, we hypothesize that the moderating effect of advertising was highly conditional on the industry and the size of the business. We expect the benefits of advertising to be amplified in industries more severely affected by the pandemic and for smaller businesses that may otherwise have fewer resources and capabilities to survive.

\section{Data and Measures}

\subsection{Measuring Business Survival}

We obtain data on business survival during the COVID-19 pandemic from the Future of Business Survey (FOBS). The FOBS is a joint data collection project conducted by the World Bank, the Organisation for Economic Co-operation and Development (OECD), and Facebook that collects survey data on the challenges, opportunities, and needs of businesses around the world using the Facebook platform. First issued in 2016, the FOBS is usually fielded twice a year in more than 90 countries; however, it was adapted in March 2020 to obtain timely data on businesses during the pandemic. In this study, we use FOBS data collected between May 20 and October 31, 2020 from more than 60,000 SMBs across 49 countries to measure business survival.

The FOBS uses a probability-based sampling design to collect data from a random sample of Facebook Business Page administrators at the country level. A Facebook Business Page functions as the organizational equivalent of a Facebook profile, serving as a business’s website in some cases, and the FOBS is designed to be representative of the population of Facebook Business Pages within each country. While FOBS does not cover all businesses in each country, sampling from this target population has several advantages. Notably, it enables rapid and timely data collection using probability sampling, which facilitates statistical inference, unlike most panel-based online surveys. It also enables sampling from the same target population across countries. Facebook Business Pages have a low barrier to adoption; nevertheless, our sample is limited to businesses with a digital presence on Facebook. We discuss this limitation and the generalizability of our results later.

We use the FOBS data to construct an outcome variable of business survival. The FOBS includes a question on whether a business is currently operational or engaging in any revenue-generating activities. We use this question to construct a dichotomous variable of whether a business had failed at the time of the survey. If a survey taker responds negatively to this question, the FOBS follows up by asking when the business was last operational. We use this second question to construct a variable for when the business failed, with the answer choices ranging from before January 1, 2020, to after May 1, 2020, with monthly intervals in between (i.e., January 1 to 31, 2020, February 1 to 29, 2020, and so on). In this way, we construct a discrete-time survival variable with business-period observations. Table \ref{table:summary} provides summary statistics for this variable and all others.

\subsection{Measuring Government Responses to the COVID-19 Pandemic}

We obtain data on government responses to the COVID-19 pandemic from the Oxford COVID-19 Government Response Tracker (OxCGRT). The OxCGRT is a data collection project that gathers information on the policies and interventions that governments around the world have adopted in response to the pandemic. Data on these policies and interventions come from publicly available sources, including news articles and government press releases, which are then used to construct 19 indicators observed on a daily basis in more than 180 countries. These indicators provide time-series cross-sectional data across three main areas: containment and closure policies, economic policies, and health system policies. 

The OxCGRT uses these indicators to construct four policy indices that measure the extent of government responses: a containment and health index, a stringency index, an economic support index, and an overall government response index.\footnote{The constituent indicators of each policy index are available in \citet{hale2020variation}.} Aggregating these indicator variables to construct a composite index introduces several limitations: it takes away the nuances of each indicator, and the interpretability of the index is limited to the indicators that are included. However, this approach is also advantageous in several ways. It mitigates the possibility of any one indicator being over- or misinterpreted, which is especially relevant when coding COVID-19-related policy responses due to subtleties and heterogeneity that exist between governments \citep{hale2020variation}, and it also allows for a measure of the overall intensity of policy responses that is comparable across countries. The OxCGRT constructs these four measures as an additive index of its constituent indicators. The stringency index corresponds to the containment and closure indicators, the economic support index corresponds to the economic policy indicators, the containment and health index corresponds to the containment and closure and health system indicators, and the overall government response index corresponds to all three categories of indicators. The appendix provides more details on the OxCGRT data.

The indices are also coded in a way to distinguish between national and subnational variation in government responses. This is vital because responses can vary substantially between levels of government. The OxCGRT handles this issue by distinguishing between government responses that encompass a geographic region and those that are implemented by specific levels of government. This process enables us to examine government responses that apply to a whole country. We focus on the overall government response index as our main variable of government lockdowns to examine the role of social and economic restrictions in the context of economic support provided by governments, though we also investigate the role of different indices in additional analyses. We construct our government response variables by aggregating the daily observations of these indices into a monthly average to match the structure of our survival variable, and we center these variables at their means to facilitate interpretation.

\subsection{Measuring Advertising Technology Use}

By advertising technology, we refer to the advertising functionalities provided by digital online platforms. This technology is characterized by large-scale reach, high levels of personalization, and an ability for the advertising firm to instantly react to strategic and tactical adjustments. We obtain data on advertising technology use from Facebook’s advertising database. Advertising on Facebook enables businesses to reach consumers who they believe will be most interested in their ads according to a variety of behavioral and demographic characteristics with highly personalized messages. A key benefit of using this data source is that it enables us to observe the actual advertising behavior for businesses in our sample as opposed to self-reported advertising behavior. We collect data on the advertising use and spend history of businesses in the FOBS sample. Using these data, we operationalize a firm's degree of advertising technology use as its expenditure on online advertising on the Facebook platform.

Studies on business survival anticipate a lagged effect of advertising \citep{shumway2001forecasting,beaver2005have, jindal2015impacts}, which is consistent with our theoretical expectations. These studies lag their advertising variable by one to three years. We follow these studies by constructing our advertising variable in three ways. First, we construct a variable of the monthly average of advertising expenditures made by each business on the Facebook platform in the year before the pandemic, which we log-transform. Measuring advertising expenditures is a common way to operationalize advertising in previous studies \citep{sethuraman2011well}. Second, we construct a dichotomous variable of whether each business in our sample engaged in online advertising on the Facebook platform in the year before the pandemic. We use this variable as a simpler measure to distinguish between advertisers and non-advertisers. Third, we construct a variable of the monthly average number of online advertisements from each business on the Facebook platform during the same period, which we also log-transform. This variable should be correlated with the amount of advertising expenditures, but it should better account for the variety of advertising activities from each business. We focus on advertising expenditures as our main variable of advertising, but we also examine the role of the other two variables in additional analyses. 

We obtain these data by referring to the Facebook profile ID of each FOBS respondent and matching these users with their Facebook advertising account. This enables us to observe the advertising behavior of each business in our sample, but it introduces an issue in which some Facebook profile IDs are associated with more than one advertising account. This can occur when a user is affiliated with more than one business, and in such cases, it is not possible to accurately identify the account to which survey responses are referring. We therefore limit our FOBS sample to businesses with a single advertising account. This ensures that each FOBS respondent is matched with the correct advertising account, though it leaves us with a larger proportion of small and midsize businesses.

\section{Method}

We use the collected data to examine how SMBs responded to government lockdowns during the COVID-19 pandemic and whether this relationship is moderated by their advertising behavior. Prior studies have modeled business closure as a binary outcome and employed hazards models to analyze the closure or failure of SMBs over time \citep{shumway2001forecasting,chava2004bankruptcy, beaver2005have, campbell2008search}. We follow these studies and estimate discrete-time hazard models to analyze business survival during the pandemic. We adopt a discrete-time approach because the underlying hazard function of our analysis is continuous as a business can fail at any point in time, but our predictors are observed at discrete time intervals. Discrete-time hazard models are commonly estimated with a logit or cloglog link function, which only differ when many observations have a probability of failure exceeding 50\% \citep{beck1998taking}. We do not observe this in our sample, and therefore we estimate our model using a logit link function for interpretability. Our data follow a nested structure such that businesses in our sample are situated within countries, and we observe country- and business-level variables over time. We therefore base our analysis on the following model:
$$P(y_{ijkt}=1) = logit^{-1} \big[ \phi Gov_{jt} + \gamma Pand_{t} + \theta (Gov_{jt} \times Pand_{t}) + \bm{\beta}\bm{x}_{i} + \rho_{j} + \nu_{k} + \alpha_{t} \big],$$
where, for business $i$ in industry $k$, country $j$, and period $t$, $y$ is a dichotomous variable of business closure, $Gov$ is a continuous variable representing the government response, $Pand$ is a dichotomous variable for when the global pandemic was occurring, $\bm{x}$ is a vector of business-level covariates, $\rho$ are country fixed effects, $\nu$ are industry fixed effects, and $\alpha$ are period fixed effects. We follow prior research that uses difference-in-differences for discrete-time hazard models and estimate $\theta$ as the effect of government responses on business survival.\footnote{\citet{chen2008effects} and \citet{li2011did} provide examples of difference-in-differences for discrete-time hazard models.} Essentially, we estimate the difference in survival probability for businesses that experienced restrictions more severely during the pandemic versus those that did not. As \citet{ai2003interaction} note, difference-in-differences in non-linear models cannot be interpreted in the same way as linear models, and they must be evaluated using the results of all other covariates in the model. We follow this procedure as an extension to our analysis.

The validity of our estimates rests on the assumption that the survival probability of businesses under stricter government lockdowns would have followed a parallel trend compared with those under less strict government lockdowns. This assumption is plausible in the context of our study for three reasons. First, the measure of government lockdowns we include in our model is implemented across an entire country, which means that these lockdowns should be uncorrelated with whether a business is more likely to survive in a country. Second, recent studies investigating government responses to the pandemic find that the timing and intensity of government lockdowns are largely explained by differences in political characteristics, including regime type, formal political institutions, and party affiliation \citep{cheibub2020rights, sebhatu2020explaining,adolph2021pandemic, baccini2021explaining}. We do not expect business survival to be systematically related to these political characteristics; nonetheless, we include country fixed effects in our model to control for these country characteristics. Third, we do not expect government lockdowns to be correlated with any other characteristics that affect business survival, such as business size, access to credit, profitability, and industry \citep{denis2007duration, jindal2015impacts, lopucki2015bankruptcy}. We therefore use this model to estimate the effect of government lockdowns on business survival during the pandemic. 

We also build on this model to examine whether advertising technology moderates the impact of government lockdowns. The central problem with this task is that advertising is likely endogenous, as there could be unobserved variables driving both advertising and business survival. We address this problem in two ways. First, we include three important control variables in our model according to the business survival literature: the size of the business, whether the business had access to credit, and the industry of the business. Prior research has found that these variables are correlated with advertising expenditures and affect business survival \citep{denis2007duration, jindal2015impacts, lopucki2015bankruptcy}. We use data from the FOBS to construct these variables and code them as categorical predictors corresponding to the answer choices of each survey question. Second, we estimate our model using instrumental variables for the advertising practices of businesses. Following the literature on advertising and business survival, we use competitor-average advertising, or the average advertising of businesses operating in the same industry, as an instrument for advertising \citep{sridhar2016relating, srinivasan2018corporate,jindal2020impact}. The logic behind this instrument is that some industry-wide norm influences a business’s tendency to advertise, but this should be uncorrelated with other business characteristics that affect survival, such as management practices or stronger performance. We obtain data for these variables using the FOBS to organize businesses into industries. We compute the average advertising expenditures for each industry in each country using Facebook’s advertising data. We then follow prior studies to construct these instruments using a control function approach since our model is non-linear \citep{luan2010forecasting, petrin2010control,albuquerque2012measuring, guo2016control}. \citet{abrevaya2010testing} argue that standard approaches for linear models, such as two-stage least squares, produce inconsistent estimates for non-linear models. When constructing these instruments, we find that both the estimates of these instruments and the F-statistic in the first-stage regressions of the control function approach are highly significant, which confirms the relevance of these instruments. In addition to using these instrumental variables in our analysis, we extend our model to include a three-way interaction term to examine how businesses' advertising practices moderate the difference-in-differences estimate. We present our estimates with cluster-robust standard errors with clustering at the country and period levels.

\section{Results}

We present our main results in Table \ref{table:main_results}. The first column shows the estimated effect of government lockdowns on business survival, and the next two columns show how online advertising moderates this effect. A negative sign on the coefficients indicates a lower risk of closure. The first column shows the results of our discrete-time survival model with country and industry fixed effects. The coefficient on the interaction term is positive and significant, which indicates that government lockdowns overall increased the likelihood of business closures around the world during the pandemic. While some research finds that government lockdowns did not adversely affect business performance, our findings are consistent with the view that stay-at-home orders in fact negatively affected business performance and made it more difficult for businesses to endure the pandemic \citep{gourinchas2020covid,alexander2020stay}. 

The second and third columns of Table \ref{table:main_results} show how the impact of government lockdowns is conditional on whether businesses engaged in online advertising. Here, the interaction term between the pandemic and government response variables remains positive. However, the three-way interaction term of the pandemic, government response, and advertising expenditure variables is negative and significant. This result suggests that while government lockdowns made surviving more difficult, businesses that engaged in online advertising were less likely to close during the pandemic. Note that these analyses are based on a control function approach to address the potential endogeneity concern of advertising and that relevant control variables along with country and industry fixed effects are included. Thus, we argue that the results are not merely a reflection of business-level characteristics that may be driving both advertising and survival during the pandemic. Our findings are also consistent with the view that advertising helps reduce bankruptcy risk even in times of economic turbulence \citep{jindal2015impacts}. 

Our results also show that larger businesses and businesses with access to credit were less likely to have closed during the pandemic. This result aligns with both theoretical and empirical research on the drivers of business failure \citep{chava2004bankruptcy, denis2007duration}. The results presented in Table \ref{table:main_results} are therefore consistent with our hypotheses thus far; however, interpretability of the estimates is limited because the model is non-linear. We therefore perform several extensions to make the results more interpretable. First, we compute the difference in survival probability for an average increase in government lockdowns that occurred during the first month of the pandemic across our full sample, holding all other variables at their mean levels. We also compute these results for different baseline levels of lockdowns and for each period of the pandemic in our study. Second, we compute the difference in survival probability between businesses that did not advertise and those with a one standard deviation increase in advertising expenditures, holding all other variables at their mean levels. We also compute these results for different levels of lockdowns and for each period.

\begin{figure}[ht]
  \begin{minipage}[t]{\textwidth}
    \centering
    \includegraphics[scale=0.7]{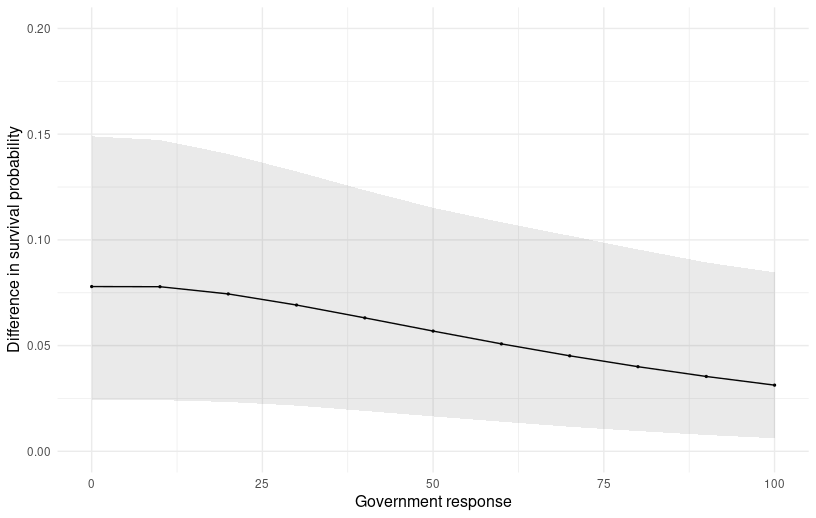}
    \caption{Estimated Effect of Lockdowns by Stringency Level}
    \label{fig:lockdown_stringency}
  \end{minipage}
  
  \vspace{1cm} 
  
  \begin{minipage}[t]{\textwidth}
    \centering
    \includegraphics[scale=0.7]{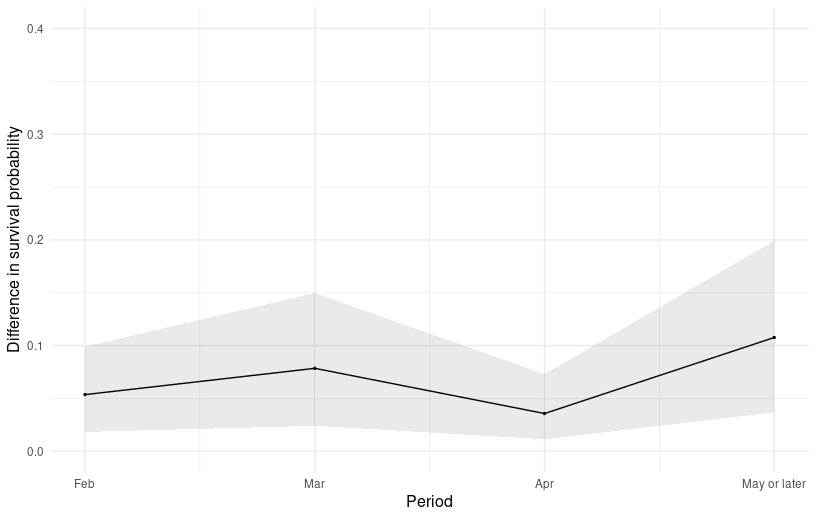}
    \caption{Estimated Effect of Lockdowns by Period}
    \label{fig:lockdown_period}
  \end{minipage}

\end{figure}

We find that an average increase in government lockdowns is associated with a 7.8 percentage point increase in the probability of closure across our full sample. Figure \ref{fig:lockdown_stringency} shows the estimated differences in the probability of closure for different baseline levels of government lockdowns. We find that the impact of these lockdowns was strongest at lower baseline levels and diminishes as these levels increase. This suggests that businesses had the most difficulty surviving the initial shock of the lockdowns, while subsequent increases in the severity of these lockdowns had less of an impact. Figure \ref{fig:lockdown_period} shows the estimated difference in the probability of closure for each period of our study. We find that the impact of lockdowns was greater in March and May or later, with a slight decrease in April. This result broadly corresponds to our findings in Figure \ref{fig:lockdown_stringency}, in which lockdowns were at or close to zero in most countries before the pandemic was declared in mid-March, increased substantially in April, and decreased in May before leveling off afterward.

\begin{figure}[ht]
  \begin{minipage}[t]{\textwidth}
    \centering
    \includegraphics[scale=0.7]{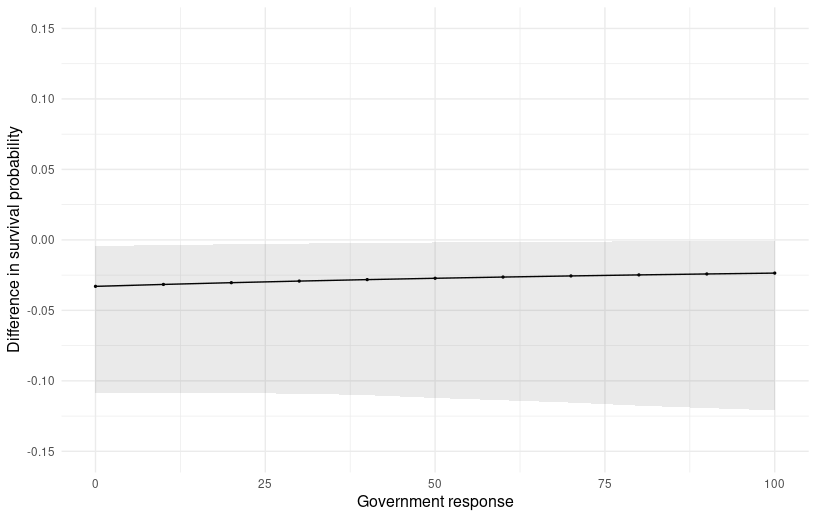}
    \caption{Estimated Effect of Advertising by Stringency Level}
    \label{fig:ad_stringency}
  \end{minipage}
  
  \vspace{1cm} 
  
  \begin{minipage}[t]{\textwidth}
    \centering
    \includegraphics[scale=0.7]{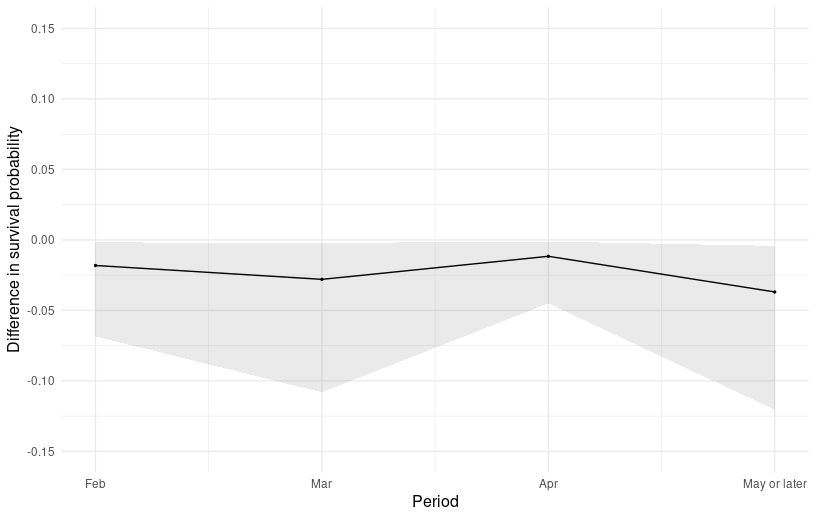}
    \caption{Estimated Effect of Advertising by Period}
    \label{fig:ad_period}
  \end{minipage}

\end{figure}

We also find that a one standard deviation increase in advertising is associated with a 2.8 percentage point decrease on average in the probability of closure across our full sample. Figure \ref{fig:ad_stringency} shows the estimated differences in the probability of closure for different levels of lockdowns. We find small changes in this estimate across different levels of lockdowns. Although the estimated differences in the probability of closure increase slightly as lockdowns become more severe, they stay close to its mean of 2.8. Figure \ref{fig:ad_period} shows the estimated difference in the probability of closure for each period of our study. Again, we find small differences in this estimate across periods, though still larger in March and May or later, which corresponds to the pattern observed in Figure \ref{fig:lockdown_period}. Overall, we find evidence that both government lockdowns and advertising expenditures had an impact on business closures during the pandemic. The results thus far, however, are based on our full sample. In the following subsections, we investigate potential heterogeneous responses that may exist between key subpopulations, including analyses by country, industry, and business size.

\subsection{Results by Country and Region}

We use the results in the third column of Table \ref{table:main_results} to calculate the estimated effect of government lockdowns across countries. Similar to our previous analysis, we compute the difference in probability of survival for an average increase in government lockdowns that occurred at the onset of the pandemic, while using the estimated fixed effects to capture variation among countries and holding all other variables at their average levels. Due to page limit, the detailed result is provided in Table B1 in Web Appendix B. The result shows substantial variation among countries, with an equivalent increase in government lockdowns having the strongest impact in India, Bangladesh, and Uganda and the weakest impact in Japan, Sweden, and Poland. The effect sizes of these estimates range from 2.3 to 15.9 percentage points, with a mean of 7.8 across countries. Although we find few studies with whom to compare these results, \citet{gourinchas2020covid} use prepandemic data to estimate an increase in the failure rate of small and medium-sized enterprises across 17 countries of approximately 9.0 percentage points due to the COVID-19 pandemic, absent government support. Their estimate is comparable to ours, though our analyses also consider the role of government support measures. Regionally, we find that countries in Asia and Africa were most susceptible to the effects of government lockdowns: the top 10 countries most severely affected were from these regions, with the United Kingdom standing out as a notable exception. Conversely, we find that countries in Europe tended to be the least susceptible to the effects of lockdowns, with eight of the bottom 10 countries being from Europe. The two non-European countries in this list are Japan and South Korea. The three North American countries in our sample (the United States, Mexico, and Canada) are close to the cross-country median in the estimated effect of government lockdowns.

Next, we examine whether online advertising moderated the impact of government lockdowns across countries. We build on the previous analysis by calculating the differences in survival probabilities of SMBs that did not advertise with those that increased their advertising expenditures by one standard deviation for each country, holding all other variables at their mean levels during the first month of the pandemic. The detailed result is provided in Table B2 in Web Appendix B. The result shows that the effect of an equivalent increase in advertising expenditures ranged from 0.87 percentage points to 6.3 percentage points, with a mean of 2.8 percentage points. While we cannot compare these estimates with prior studies because few have examined the effect of advertising on business survival, in their study of more than 9,000 US firms from 1980 to 2006, \citet{jindal2015impacts} find that a 1\% increase in advertising assets was associated with a 0.04\% to 0.22\% lower risk of entering bankruptcy, depending on the level of market turbulence. By contrast, when we reinterpret our results, we find that a 1\% increase in online advertising is associated with a 0.03\% decrease in the probability of business closure across our full sample of countries during the pandemic, while holding government lockdowns at a minimal level. Regionally, we find that the countries where online advertising mattered most are in Asia, with seven of the top 10 countries being from this region, including India, Israel, Bangladesh, and Malaysia. The three exceptions in this list are the United Kingdom, Argentina, and Australia. The countries where advertising appears to have mattered the least are also in Asia and Europe, with nine of the bottom 10 countries being from these regions and the estimates from Japan, Poland, South Korea, and Sweden being the smallest. The one exception in this group is Mexico. The other two North American countries, Canada and the United States, are in the top half of our sample of countries. We therefore find broad evidence to support our hypotheses, but we also discover substantial variation in the effect of both advertising and government lockdowns across countries.

\subsection{Results by Industry and Size}

We extend our analysis to examine heterogeneous effects that might exist by industry and businesses size. Recent studies have found that SMBs were especially susceptible to failure during the pandemic, and certain industries, such as the accommodations, hospitality, and arts and entertainment industries, were the most severely affected \citep{apedo2020unmasking, gourinchas2020covid}. We therefore repeat the preceding analysis to uncover any differences that might exist between these groups. Tables \ref{table:lockdown_by_industry} and \ref{table:lockdown_by_size} report the results. First, we find that businesses in the services industry (e.g., performing arts and entertainment, education and childcare services, hospitality and events services) were the most severely affected, followed by the transportation and logistics and hotels, cafes, and accommodations industries. This is largely consistent with the industries that were found to be heavily affected by stay-at-home orders, and a commensurate increase in government lockdowns decreased the survival probability of businesses in these industries by 7.9 to 8.7 percentage points. Second, businesses in the agricultural, construction, and manufacturing industries were the least affected, with an increase in lockdowns reducing the survival probability of businesses in these industries by 5.9 to 6.4 percentage points. Regarding the effects of lockdowns by business size, we find that smaller businesses were more likely to have been negatively affected than larger businesses. An equivalent increase in government lockdowns had an impact on the smallest businesses that was more than 1.5 times higher than that of the largest businesses, with the decrease in survival probability ranging from 5.2 percentage points to 8.1 percentage points.

Regarding the effect of advertising by industry, Table \ref{table:ad_by_industry} shows that the services, information and communication, and hospitality industries obtained the greatest benefit from advertising expenditures. The difference in effect size between these industries was small, ranging from 2.7 to 3.1 percentage points. Businesses in the agriculture, construction, and manufacturing industries attained the smallest benefit from advertising, with the probability of survival increasing by approximately 2 percentage points for the same increase in advertising expenditures. Regarding the effect of online advertising by business size, Table \ref{table:ad_by_size} shows substantial variation between smaller and larger businesses. We find that SMBs with fewer than 10 people had an increase in survival probability that was more than 1.8 times that of SMBs with 500 employees or more. These results indicate that small businesses not only were affected more severely by government lockdowns but also benefited more from advertising. Overall, we find compelling evidence that government lockdowns during the pandemic adversely affected business survival around the world but that online advertising helped mitigate this impact. These findings are consistent with our hypotheses, and we also uncover notable patterns across industries and business sizes. Next, we check the robustness of our results with several additional analyses. 

\section{Additional Analyses}

\subsection{Alternative Measures of Advertising}

In our main analysis, we measure online advertising technology use as the average monthly online advertising expenditures for each business. Research commonly uses advertising expenditures to operationalize advertising for empirical analysis \citep{sethuraman2011well}. However, we rerun our analysis with the two other measures of advertising (i.e., advertiser status and number of advertisements) to test whether our results are sensitive to the inclusion of these variables. First, we fit our model using our measure of advertiser status. According to our hypotheses, we expect a difference in survival between advertisers and non-advertisers, regardless of the level of advertising expenditures. Second, we fit our model with our variable for the average number of advertisements from each business. We expect the number of advertisements to be correlated with advertising expenditures, but it should better reflect the effect of having a wider variety of advertisements. Table \ref{table:alternative_ad_metrics} provides the results. After running these models, we find that our results largely mirror those in Table \ref{table:main_results}. The estimate of the three-way interaction term is negative and significant in both scenarios; thus, the results are robust to these alternative ways of measuring online advertising. 

\subsection{Alternative Measures of Government Lockdowns}

In our main results, we measure government lockdowns using the OxCGRT’s government response index, which includes the containment and closure indicators, economic policy indicators, and health system indicators. Here, we rerun our analysis using two alternative indices: the stringency index and the containment and health index. The stringency index includes only the containment and closure indicators, and the containment and health index includes both the containment and closure indicators and health system indicators. Both indices capture lockdown measures through the containment and closure indicators, but they do not include the government support measures from the economic policy indicators. We examine whether our results are sensitive to the use of either of these alternative indices. We expect these indices to produce similar results to those of the government response index, with the stringency index likely to have the greatest negative impact on survival. As Table \ref{table:alternative_lockdown_measures} shows, the results from this analysis are substantively consistent with our main results, with only minimal changes in our key estimates.

\subsection{Additional Control Variables}

Previously, we explained how our method addresses the issue of confounding variables. Here, we consider several additional variables that may be pertinent to our analysis. Recent studies on the COVID-19 pandemic have found that the fear of contracting COVID-19  was an important factor that led people to voluntarily stay at home, which adversely affected businesses no matter whether governments implemented lockdowns \citep{chetty2020real, goolsbee2021fear, sears2023we}. We therefore rerun our analysis by including several variables in an effort to control for the intensity of the pandemic in each country including the total and new number of COVID-19 cases for each country per period, which we obtain from Oxford's COVID-19 database and log-transform \citep{owidcoronavirus}. We also consider the role of several political and macroeconomic variables. Recent studies have underscored the role of political characteristics in explaining the variation in government responses to the pandemic \citep{cheibub2020rights,sebhatu2020explaining,adolph2021pandemic,baccini2021explaining}. Although we do not expect these characteristics to be systematically related to business survival, we rerun our analysis while controlling for several of these characteristics. We include the Voice and Accountability indicator and Government Effectiveness indicator from the World Bank Governance Indicators as measures of regime type and the quality of public and civil services, respectively. Finally, we control for income per capita as a measure of economic development, which we obtain from the World Bank's World Development Indicators and log-transform. As Table \ref{table:additional_controls} shows, our results are robust to the inclusion of these variables, with minor changes in our two main coefficients of interest.

\subsection{Alternative Time Frames}

Next, we investigate whether our results are dependent on the time frame used in our analysis. For our main results, we estimate the effect of advertising expenditures made in the year before the pandemic. This is because we expect advertising expenditures to have a lagged effect on the performance and survival of businesses, and advertising that occurred during or shortly before the pandemic may not have had sufficient time to affect business outcomes. Nonetheless, we alter our time frame to test whether our results are sensitive to this change. First, we construct another version of our advertising variable by calculating the average monthly advertising expenditures for each business during the first half of 2019. Doing so allows us to increase the lag between our advertising variable and the beginning of the pandemic, which serves as a more rigorous test of a lagged effect. Second, we reconstruct this variable by examining the effect of advertising expenditures just before and during the pandemic. Our theoretical approach aligns with the view that advertising produces market-based intangible assets that persuade consumers to purchase from businesses, thereby increasing sales in future periods. Most empirical studies in the marketing literature provide evidence that is consistent with this persuasion view of advertising; however, this view also suggests that advertising should be able to affect business sales more immediately \citep{joshi2010direct}. We therefore rerun our analysis to determine whether this is true. Doing so enables us to verify the information view of advertising, which suggests that advertising mostly affects current sales \citep{mcalister2016advertising}. We construct our advertising variable by taking the log of monthly online advertising expenditures for each business two months before each period of study. For example, we examine whether advertising expenditures for each business in January and February 2020 moderated the effect of government lockdowns and business survival in March 2020. Table \ref{table:alternative_timeframes} shows that our main results continue to hold after increasing the lag between advertising expenditures and the pandemic. Regarding the impact of advertising expenditures in the previous two months of each period, we also find that our estimates on the interaction terms retain their direction and significance, though the magnitude of the estimate decreases slightly. 

\subsection{Accounting for Staggered Policy Implementation}

In the main analysis, we use a discrete-time hazard model to analyze the effect of the pandemic and government lockdown on business survival through a difference-in-differences setup. Note that different countries implemented lockdowns at different points in time, causing the ``shock'' to businesses to occur in different periods. Recent research on difference-in-differences has found that the standard two-way fixed effects (TWFE) approach as adopted in the main analysis might be biased when shocks occurred in a staggered manner \citep{goodman-baconDifferenceindifferencesVariationTreatment2021}. We address this concern by adopting an extended two-way fixed-effect (ETWFE) approach \citep{wooldridgeTwoWayFixedEffects2021, wooldridgeSimpleApproachesNonlinear2022, bermanValueDescriptiveAnalytics2022}. Intuitively, the ETWFE approach extends the TWFE approach by adding additional interaction terms, including those between treatment and periods, as well as between treatment and cohorts, where each cohort includes countries that adopted the treatment in the same period. The advantage of ETWFE is that it addresses the staggered adoption concern and can be applied to non-linear models.

We specify two models and estimate both using the ETWFE approach. The first approach is a linear probability model (LPM), where the outcome variable is whether a business is closed in a given period. The second approach analyzes the same outcome variable but instead uses a logit model. In both ETWFE models, we explicitly account for the different periods in which the countries adopted lockdown policies. We use a binary variable (i.e., whether a stay-at-home order was in effect) as the treatment in both the LPM and logit models. In addition, the ETWFE approach enables testing of heterogeneous treatment effects, and therefore we further break down the analysis by industry.

Tables \ref{table:staggered_lpm} and \ref{table:staggered_logit} report the results of the LPM and logit models, respectively. We find that certain industries (e.g., services, hotels, transportation) were affected more by a government's lockdown policy (i.e., the stay-at-home order). Importantly, as shown in Column (3) of both tables, we find that businesses that did not advertise on Facebook had significant increases in business closure probability across different industries in both models, whereas businesses that did advertise on Facebook (i.e., Column (2)) did not show any significant increase. Overall, the results are largely consistent with the main analysis. They corroborate that online advertising attenuated the adverse effects of governmental lockdown policies on business survival across countries and business sectors.

\section{Conclusion}

In this study, we set out to examine whether government lockdowns affected the survival of SMBs during the COVID-19 pandemic and to investigate the extent to which use of online advertising technology moderated this effect. Recent studies have examined how the COVID-19 pandemic has shaped the performance and survival of businesses, but to date, whether and how advertising technology may have helped businesses navigate related demand shocks is unclear. Drawing from the theoretical and empirical research on advertising and the recent literature on business survival during the pandemic, we hypothesize that stay-at-home orders increased business closures but that online advertising expenditures helped mitigate this impact. We also hypothesize that this relationship was conditional on industry and business size, with the benefits of advertising amplified in industries most severely affected by the pandemic and for smaller businesses with fewer resources and capabilities to survive. 

Using data from more than 60,000 businesses in 49 countries collected on the Facebook platform from May to October 2020, we test our hypotheses using discrete-time survival models with instrumental variables. Our analyses uncover a compelling set of findings, specifically that government lockdowns led to a decrease in survival probability of businesses around the world and that advertising expenditures helped attenuate this effect. We also find substantial variation in the results among regions, countries, industries, and business sizes. Businesses in Africa and Asia (with the exception of Japan and South Korea) seem to have been more susceptible to the adverse effects of stay-at-home orders, while businesses in Europe appear less susceptible. Small businesses with fewer than 10 employees were the most severely affected by lockdowns, but they also had the greatest gains from advertising. In addition, businesses in the services, transportation, and hospitality industries were the most severely affected, while businesses in most of these industries also attained the greatest benefits from advertising. Larger SMBs with 500 or more employees and businesses in the agricultural, manufacturing, and construction industries were the least adversely affected by government lockdowns and obtained the lowest gains from online advertising. 

Our findings extend the literature in several ways. First, despite the recent growth of research on the effects of the COVID-19 pandemic, researchers are divided on whether government lockdowns negatively affected businesses during the pandemic. Most of this research is focused on the experiences of businesses in the United States. In this study, we analyze an original and unique data set and uncover evidence that lockdowns indeed intensified business closures around the world \citep{gourinchas2020covid,alexander2020stay}. Second, we extend the literature by investigating whether and how advertising technology was a useful tool for SMBs during the pandemic. Prior research finds that advertising in general helps US businesses survive during times of economic turbulence \citep{jindal2015impacts,angulo2022marketingfundingrecession}, but the impact of online advertising in particular on business survival is not well understood. Our findings indicate that online advertising technology played an important role in alleviating some of the adversity businesses faced during the pandemic. These findings are all the more important as reports suggest that firms historically tended to cut back rather than increase marketing investments in times of severe economic turmoil \citep{oezturan2014downturnturkey,rollins2014downturnmarketing}. Third, our findings are consistent with the view that advertising improves business performance through a lagged effect. This finding has noteworthy implications for businesses. Our study suggests that advertising technology not only is beneficial in improving revenues and repeat purchases for businesses \citep{manchanda2006effect, lewis2014online} but also serves as an ``insurance'' against sudden and extreme demand shocks from global events.

Our study also has several limitations and raises questions that have yet to be answered. A first question is whether our findings apply to other modes of advertising. The predictions of our theoretical framework are not limited to the use of personalized online advertising, and we expect that different modes of advertising, such as print and television advertising, might also have supported the performance and survival of businesses during the pandemic. However, evidence on whether these other modes of advertising mattered in the same way is limited. A second question is whether online advertising technology was especially relevant during the pandemic in light of the increase in internet usage and e-commerce expansion. More fully understanding how different modes of advertising interacted with the increased internet use during the pandemic is important. While our study focuses on SMBs, which likely used online advertising technology as their primary advertising medium, we do not observe these businesses' advertising expenditure across other media. Gaining insights into their overall advertising spending and allocation across different media and channels would shed further light on the role of online advertising in enhancing business resilience -- including potential interaction effects between different media. A third question is how spending on advertising technology may have interacted with further strategic adjustments and investments that in turn may have been easier for specific businesses across and within industries \citep{noel2022hotelcovid}. A fourth question may pertain to credit access and small business lending practices. Future research could study if and how banks and other institutions could support SMBs' spending on advertising technology usage with loans during severe exogenous demand shocks \citep{alekseev2023effects}.

\vspace{15mm}

\subsection*{Funding and Competing Interests}
Author B was employed by Company A (before renaming of Company A to Company B). Author B further worked as an independent consultant for Company B. Both engagements were fully unrelated to the research work for the submitted manuscript. Authors C and E are employed by Company B. Author F was employed by Company B. Authors C, E, and F own Company B stocks. Authors A and D have no affiliations with or involvement in any organization or entity with any financial interest or non-financial interest in the subject matter or materials discussed in this manuscript.
 
Company B has a right to review and require changes to the paper to ensure compliance with Company B’s privacy policy and legal commitments.

\vspace{15mm}

\bibliography{covidSurvival}

\vspace{30mm}

\begin{table}[ht]
\centering
\small
\caption{Summary Statistics} 
\begin{tabular}{lrrrr}
  \hline
Variable & Min & Max & Mean & SD \\ 
  \hline
Business closure & 0.00 & 1.00 & 0.18 & 0.38 \\ 
  Business size & 1.00 & 5.00 & 1.26 & 0.71 \\ 
  Access to credit & 0.00 & 1.00 & 0.31 & 0.46 \\ 
  Advertising expenditures & 0.00 & 17.77 & 4.83 & 3.98 \\ 
  Advertiser status & 0.00 & 1.00 & 0.64 & 0.48 \\ 
  Number of advertisements & 0.00 & 7.08 & 0.75 & 0.82 \\ 
  Government response & -30.45 & 61.69 & 0.00 & 29.88 \\ 
  Stringency & -32.41 & 67.59 & 0.00 & 32.58 \\ 
  Containment health & -31.42 & 63.58 & 0.00 & 30.26 \\ 
  New deaths & 0.00 & 7.61 & 1.97 & 2.23 \\ 
  New cases & 0.00 & 10.30 & 3.02 & 3.13 \\ 
  GDP per capita & 6.68 & 11.31 & 9.67 & 1.17 \\ 
  Voice and accountability & -1.62 & 1.69 & 0.49 & 0.86 \\ 
  Government effectiveness & -1.09 & 2.22 & 0.69 & 0.80 \\ 
   \hline
\end{tabular}
\label{table:summary}
\end{table}

\begin{table}[!htbp] \centering 
\small
  \caption{Government Lockdowns, Advertising Expenditures, and Business Closure} 
  \label{} 
\begin{tabular}{@{\extracolsep{5pt}}lccc} 
\\[-1.8ex]\hline 
\hline \\[-1.8ex] 
 & \multicolumn{3}{c}{\textit{Dependent variable:}} \\ 
\cline{2-4} 
\\[-1.8ex] & \multicolumn{3}{c}{Business closure} \\ 
\\[-1.8ex] & (1) & (2) & (3)\\ 
\hline \\[-1.8ex] 
 Pandemic & 1.230$^{***}$ & 1.133$^{***}$ & 1.135$^{***}$ \\ 
  & (0.210) & (0.203) & (0.203) \\ 
  & & & \\ 
 Gov. response & $-$0.026$^{***}$ & $-$0.032$^{***}$ & $-$0.032$^{***}$ \\ 
  & (0.005) & (0.006) & (0.006) \\ 
  & & & \\ 
 Advertising expenditures &  & $-$1.181$^{***}$ & $-$0.072 \\ 
  &  & (0.254) & (0.195) \\ 
  & & & \\ 
 Size &  &  & $-$0.109$^{***}$ \\ 
  &  &  & (0.026) \\ 
  & & & \\ 
 Access to credit &  &  & $-$0.646$^{***}$ \\ 
  &  &  & (0.082) \\ 
  & & & \\ 
 Pandemic x Gov. response & 0.011$^{**}$ & 0.023$^{**}$ & 0.023$^{**}$ \\ 
  & (0.004) & (0.008) & (0.008) \\ 
  & & & \\ 
 Advertising expenditures x Gov. response &  & 0.002$^{***}$ & 0.002$^{***}$ \\ 
  &  & (0.0003) & (0.0003) \\ 
  & & & \\ 
 Advertising expenditures x Pandemic &  & 0.001 & 0.001 \\ 
  &  & (0.006) & (0.006) \\ 
  & & & \\ 
 Pandemic x Gov. response x Advertising expenditures &  & $-$0.003$^{***}$ & $-$0.003$^{***}$ \\ 
  &  & (0.0005) & (0.0005) \\ 
  & & & \\ 
\hline \\[-1.8ex] 
Period dummies & Yes & Yes & Yes \\ 
Country dummies & Yes & Yes & Yes \\ 
Industry dummies & Yes & Yes & Yes \\ 
Control function & No & Yes & Yes \\ 
Countries & 49 & 49 & 49 \\ 
Businesses & 64,155 & 64,155 & 64,155 \\ 
\hline 
\hline \\[-1.8ex] 
\multicolumn{4}{l}{Cluster-robust standard errors are in parentheses. $^{*}$p$<$0.05; $^{**}$p$<$0.01; $^{***}$p$<$0.001} \\ 
\end{tabular}
\label{table:main_results}
\end{table}

\begin{table}[ht]
\centering
\small
\caption{Estimated Effect of Government Lockdowns by Industry} 
\begin{tabular}{lrrr}
  \hline
Industry & Estimate & Lower & Upper \\ 
  \hline
  Services & 0.0865 & 0.0316 & 0.1490 \\ 
  Transportation and logistics & 0.0808 & 0.0274 & 0.1451 \\ 
  Hotels, cafes, and restaurants & 0.0785 & 0.0294 & 0.1359 \\ 
  Retail and wholesale & 0.0695 & 0.0151 & 0.1611 \\ 
  Information and communication & 0.0685 & 0.0200 & 0.1347 \\ 
  Manufacturing & 0.0636 & 0.0187 & 0.1248 \\ 
  Construction & 0.0634 & 0.0184 & 0.1263 \\ 
  Agriculture & 0.0594 & 0.0171 & 0.1163 \\ 
   \hline
   \multicolumn{4}{l}{\footnotesize{Estimates are calculated from the results of model 3 in Table \ref{table:main_results}}} \\ 
\end{tabular}
\label{table:lockdown_by_industry}
\end{table}

\begin{table}[ht]
\centering
\small
\caption{Estimated Effect of Government Lockdowns by Business Size} 
\begin{tabular}{lrrr}
  \hline
Size & Estimate & Lower & Upper \\ 
  \hline
Fewer than 10 people & 0.0806 & 0.0244 & 0.1539 \\ 
  10 to 49 people & 0.0690 & 0.0225 & 0.1264 \\ 
  50 to 249 people & 0.0611 & 0.0205 & 0.1100 \\ 
  250 to 499 people & 0.0546 & 0.0184 & 0.0978 \\ 
  500 people or more & 0.0520 & 0.0176 & 0.0924 \\ 
   \hline
   \multicolumn{4}{l}{\footnotesize{Estimates are calculated from the results of model 3}} \\ 
   \multicolumn{4}{l}{\footnotesize{in Table \ref{table:main_results}.}} \\ 
\end{tabular}
\label{table:lockdown_by_size}
\end{table}

\begin{table}[ht]
\centering
\small
\caption{Estimated Effect of Advertising by Industry} 
\begin{tabular}{lrrr}
  \hline
Industry & Estimate & Lower & Upper \\ 
  \hline
  Services & -0.0312 & -0.1136 & -0.0026 \\ 
  Information and communication & -0.0282 & -0.1239 & -0.0012 \\ 
  Hotels, cafes, and restaurants & -0.0269 & -0.0934 & -0.0028 \\ 
  Transportation and logistics & -0.0255 & -0.0847 & -0.0031 \\ 
  Retail and wholesale & -0.0251 & -0.1052 & -0.0013 \\ 
  Manufacturing & -0.0216 & -0.0824 & -0.0018 \\ 
  Construction & -0.0216 & -0.0822 & -0.0018 \\ 
  Agriculture & -0.0195 & -0.0726 & -0.0018 \\ 
   \hline
   \multicolumn{4}{l}{\footnotesize{Estimates are calculated from the results of model 3 in Table \ref{table:main_results}.}} \\ 
\end{tabular}
\label{table:ad_by_industry}
\end{table}

\begin{table}[ht]
\centering
\small
\caption{Estimated Effect of Advertising by Business Size} 
\begin{tabular}{lrrr}
  \hline
Size & Estimate & Lower & Upper \\ 
  \hline
Fewer than 10 people & -0.0290 & -0.1117 & -0.0024 \\ 
  10 to 49 people & -0.0265 & -0.1109 & -0.0017 \\ 
  50 to 249 people & -0.0226 & -0.0927 & -0.0016 \\ 
  250 to 499 people & -0.0189 & -0.0735 & -0.0017 \\ 
  500 people or more & -0.0157 & -0.0537 & -0.0021 \\ 
   \hline
   \multicolumn{4}{l}{\footnotesize{Estimates are calculated from the results of model 3}} \\ 
   \multicolumn{4}{l}{\footnotesize{in Table \ref{table:main_results}.}} \\ 
\end{tabular}
\label{table:ad_by_size}
\end{table}


\begin{table}[!htbp] \centering 
\small
  \caption{Government Lockdowns, Advertising Metrics, and Business Closures} 
  \label{} 
\begin{tabular}{@{\extracolsep{5pt}}lcccc} 
\\[-1.8ex]\hline 
\hline \\[-1.8ex] 
 & \multicolumn{4}{c}{\textit{Dependent Variable: Business Closure}} \\ 
\cline{2-5}  \\[-1.8ex]
& \multicolumn{4}{c}{\textit{Advertising Metric:}} \\ 
 
\\[-1.8ex] & \multicolumn{2}{c}{Advertiser Status} & \multicolumn{2}{c}{Number of Advertisements} \\ 
\\[-1.8ex] & (1) & (2) & (3) & (4) \\ 
\hline \\[-1.8ex] 
 Pandemic & 1.139$^{***}$ & 1.180$^{***}$ & 1.158$^{***}$ & 1.157$^{***}$ \\ 
  & (0.201) & (0.204) & (0.199) & (0.199) \\ 
  & & & & \\ 
 Gov. response & $-$0.031$^{***}$ & $-$0.032$^{***}$ & $-$0.028$^{***}$ & $-$0.028$^{***}$ \\ 
  & (0.006) & (0.006) & (0.005) & (0.005) \\ 
  & & & & \\ 
 Advertising & $-$1.454$^{***}$ & 0.222 & $-$7.343$^{***}$ & $-$2.291$^{***}$ \\ 
  & (0.222) & (0.117) & (1.452) & (0.407) \\ 
  & & & & \\ 
 Size &  & $-$0.117$^{***}$ &  & $-$0.071$^{*}$ \\ 
  &  & (0.032) &  & (0.033) \\ 
  & & & & \\ 
 Access to credit &  & $-$0.685$^{***}$ &  & $-$0.500$^{***}$ \\ 
  &  & (0.144) &  & (0.148) \\ 
  & & & & \\ 
 Pandemic x Gov. response & 0.023$^{**}$ & 0.023$^{**}$ & 0.017$^{*}$ & 0.017$^{*}$ \\ 
  & (0.008) & (0.008) & (0.007) & (0.007) \\ 
  & & & & \\ 
 Advertising x Gov. response & 0.013$^{***}$ & 0.014$^{***}$ & 0.008$^{***}$ & 0.007$^{***}$ \\ 
  & (0.002) & (0.002) & (0.002) & (0.002) \\ 
  & & & & \\ 
 Advertising x Pandemic & $-$0.055 & $-$0.078 & $-$0.028 & $-$0.026 \\ 
  & (0.050) & (0.047) & (0.033) & (0.033) \\ 
  & & & & \\ 
 Pandemic x Gov. response x Advertising & $-$0.026$^{***}$ & $-$0.027$^{***}$ & $-$0.014$^{***}$ & $-$0.014$^{***}$ \\ 
  & (0.004) & (0.004) & (0.002) & (0.002) \\ 
  & & & & \\ 
\hline \\[-1.8ex] 
Period dummies & Yes & Yes & Yes & Yes \\ 
Country dummies & Yes & Yes & Yes & Yes \\ 
Industry dummies & Yes & Yes & Yes & Yes \\ 
Control function & Yes & Yes & Yes & Yes \\ 
Countries & 49 & 49 & 49 & 49 \\ 
Businesses & 66,643 & 66,643 & 66,571 & 66,571 \\ 
\hline 
\hline \\[-1.8ex] 
\multicolumn{5}{l}{Cluster-robust standard errors are in parentheses. $^{*}$p$<$0.05; $^{**}$p$<$0.01; $^{***}$p$<$0.001} \\ 
\end{tabular} 
\label{table:alternative_ad_metrics}
\end{table}


\begin{table}[!htbp] \centering 
\small
  \caption{Government Lockdown Measure, Advertising Expenditures, and Business Closures} 
  \label{} 
\begin{tabular}{@{\extracolsep{5pt}}lcccc} 
\\[-1.8ex]\hline 
\hline \\[-1.8ex] 
 & \multicolumn{4}{c}{\textit{Dependent Variable: Business Closure}} \\ 
\cline{2-5}  \\[-1.8ex]
& \multicolumn{4}{c}{\textit{Government Lockdown Measure:}} \\
\\[-1.8ex] & \multicolumn{2}{c}{Stringency} & \multicolumn{2}{c}{Containment Health} \\ 
\\[-1.8ex] & (1) & (2) & (3) & (4) \\ 
\hline \\[-1.8ex] 
 Pandemic & 1.293$^{***}$ & 1.293$^{***}$ & 1.051$^{***}$ & 1.052$^{***}$ \\ 
  & (0.249) & (0.249) & (0.195) & (0.196) \\ 
  & & & & \\ 
 Lockdown & $-$0.031$^{***}$ & $-$0.031$^{***}$ & $-$0.026$^{***}$ & $-$0.026$^{***}$ \\ 
  & (0.007) & (0.007) & (0.004) & (0.004) \\ 
  & & & & \\ 
 Advertising expenditures & $-$1.169$^{***}$ & $-$0.060 & $-$1.181$^{***}$ & $-$0.058 \\ 
  & (0.256) & (0.198) & (0.252) & (0.198) \\ 
  & & & & \\ 
 Size & & $-$0.109$^{***}$ & & $-$0.107$^{***}$ \\ 
  & & (0.026) & & (0.025) \\ 
  & & & & \\ 
 Access to credit & & $-$0.646$^{***}$ & & $-$0.656$^{***}$ \\ 
  & & (0.081) & & (0.088) \\ 
  & & & & \\ 
 Pandemic x Lockdown & 0.027$^{*}$ & 0.027$^{*}$ & 0.019$^{*}$ & 0.019$^{*}$ \\ 
  & (0.011) & (0.011) & (0.008) & (0.008) \\ 
  & & & & \\ 
 Advertising expenditures x Lockdown & 0.002$^{***}$ & 0.002$^{***}$ & 0.001$^{***}$ & 0.001$^{***}$ \\ 
  & (0.0003) & (0.0003) & (0.0003) & (0.0003) \\ 
  & & & & \\ 
 Advertising expenditures x Pandemic & $-$0.015 & $-$0.014 & 0.010 & 0.010 \\ 
  & (0.008) & (0.008) & (0.006) & (0.006) \\ 
  & & & & \\ 
 Pandemic x Lockdown x Advertising expenditures & $-$0.004$^{***}$ & $-$0.004$^{***}$ & $-$0.003$^{***}$ & $-$0.003$^{***}$ \\ 
  & (0.0004) & (0.0004) & (0.0005) & (0.0005) \\ 
  & & & & \\ 
\hline \\[-1.8ex] 
Period dummies & Yes & Yes & Yes & Yes \\ 
Country dummies & Yes & Yes & Yes & Yes \\ 
Industry dummies & Yes & Yes & Yes & Yes \\ 
Control function & Yes & Yes & Yes & Yes \\ 
Countries & 49 & 49 & 49 & 49 \\ 
Businesses & 64,155 & 64,155 & 64,155 & 64,155 \\ 
\hline 
\hline \\[-1.8ex] 
\multicolumn{5}{l}{Cluster-robust standard errors are in parentheses. $^{*}$p$<$0.05; $^{**}$p$<$0.01; $^{***}$p$<$0.001} \\ 
\end{tabular}
\label{table:alternative_lockdown_measures}
\end{table}

\begin{table}[!htbp] \centering 
\scriptsize
  \caption{Government Lockdowns, Advertising Expenditures, and Business Closures (Additional Controls)} 
  \label{} 
\begin{tabular}{@{\extracolsep{5pt}}lccccc} 
\\[-1.8ex]\hline 
\hline \\[-1.8ex] 
 & \multicolumn{5}{c}{\textit{Dependent variable:}} \\ 
\cline{2-6} 
\\[-1.8ex] & \multicolumn{5}{c}{Business closure} \\ 
\\[-1.8ex] & (1) & (2) & (3) & (4) & (5)\\ 
\hline \\[-1.8ex] 
 New cases & $-$0.035 &  &  &  &  \\ 
  & (0.035) &  &  &  &  \\ 
  & & & & & \\ 
 New deaths &  & $-$0.063 &  &  &  \\ 
  &  & (0.059) &  &  &  \\ 
  & & & & & \\ 
 GDP per capita &  &  & 0.018 &  &  \\ 
  &  &  & (0.081) &  &  \\ 
  & & & & & \\ 
 Voice and accountability &  &  &  & $-$0.017 &  \\ 
  &  &  &  & (0.090) &  \\ 
  & & & & & \\ 
 Government effectiveness &  &  &  &  & 0.030 \\ 
  &  &  &  &  & (0.160) \\ 
  & & & & & \\ 
 Pandemic x Gov. response & 0.029$^{**}$ & 0.035$^{**}$ & 0.023$^{**}$ & 0.022$^{**}$ & 0.022$^{**}$ \\ 
  & (0.009) & (0.013) & (0.008) & (0.008) & (0.008) \\ 
  & & & & & \\ 
 Pandemic x Gov. response x Advertising expenditures & $-$0.004$^{***}$ & $-$0.005$^{***}$ & $-$0.003$^{***}$ & $-$0.004$^{***}$ & $-$0.004$^{***}$ \\ 
  & (0.0005) & (0.0004) & (0.0005) & (0.0005) & (0.0005) \\ 
  & & & & & \\ 
\hline \\[-1.8ex] 
Period dummies & Yes & Yes & Yes & Yes & Yes \\ 
Country dummies & Yes & Yes & Yes & Yes & Yes \\ 
Industry dummies & Yes & Yes & Yes & Yes & Yes \\ 
Control function & Yes & Yes & Yes & Yes & Yes \\ 
Countries & 49 & 49 & 49 & 49 & 49 \\ 
Businesses & 64,155 & 64,155 & 64,155 & 62,411 & 62,411 \\ 
\hline 
\hline \\[-1.8ex] 
\multicolumn{6}{l}{Cluster-robust standard errors are in parentheses. $^{*}$p$<$0.05; $^{**}$p$<$0.01; $^{***}$p$<$0.001} \\ 
\end{tabular} 
\label{table:additional_controls}
\end{table}

\begin{table}[!htbp] \centering 
\small
  \caption{Government Lockdowns, Advertising Expenditures, and Business Closure (Alternative Advertising Time Frames)} 
  \label{} 
\begin{tabular}{@{\extracolsep{5pt}}lcc} 
\\[-1.8ex]\hline 
\hline \\[-1.8ex] 
 & \multicolumn{2}{c}{\textit{Dependent variable:}} \\ 
\cline{2-3} 
\\[-1.8ex] & \multicolumn{2}{c}{Business closure} \\ 
\\[-1.8ex] & (1) & (2)\\ 
\hline \\[-1.8ex] 
 Pandemic & 1.135$^{***}$ & 1.157$^{***}$ \\ 
  & (0.205) & (0.214) \\ 
  & & \\ 
 Gov. response & $-$0.030$^{***}$ & $-$0.026$^{***}$ \\ 
  & (0.006) & (0.006) \\ 
  & & \\ 
 Advertising expenditures & $-$0.066 &  0.093\\ 
  & (0.204) &  (0.052)\\ 
  & & \\ 
 Size & $-$0.103$^{***}$ & $-$0.121$^{***}$ \\ 
  & (0.024) & (0.031) \\ 
  & & \\ 
 Access to credit & $-$0.626$^{***}$ & $-$0.720$^{***}$ \\ 
  & (0.082) & (0.122) \\ 
  & & \\ 
 Pandemic x Gov. response & 0.020$^{**}$ & 0.014$^{*}$ \\ 
  & (0.007) & (0.006) \\ 
  & & \\ 
 Advertising expenditures x Gov. response & 0.002$^{***}$ &  0.001$^{***}$\\ 
  & (0.0004) &  (0.0004)\\ 
  & & \\ 
 Advertising expenditures x Pandemic & $-$0.011 & 0.015$^{*}$ \\ 
  & (0.007) & (0.007) \\ 
  & & \\ 
 Pandemic x Gov. response x Advertising expenditures & $-$0.004$^{***}$ & $-$0.003$^{***}$ \\ 
  & (0.001) & (0.0004) \\ 
  & & \\ 
\hline \\[-1.8ex] 
Period dummies & Yes & Yes \\ 
Country dummies & Yes & Yes \\ 
Industry dummies & Yes & Yes \\ 
Control function & Yes & Yes \\ 
Countries & 54 & 54 \\ 
Businesses & 64,786 & 62,949 \\ 
\hline 
\hline \\[-1.8ex] 
\multicolumn{3}{p{0.75\linewidth}}{

Note: Column (1) reports results using the advertising expenditure during the first half of 2019. Column (2) reports results using the advertising expenditure two months before each period of study. Cluster-robust standard errors are in parentheses. $^{*}$p$<$0.05; $^{**}$p$<$0.01; $^{***}$p$<$0.001
} \\ 
\end{tabular}
\label{table:alternative_timeframes}
\end{table}

\begin{table}

\caption{Effect of Stay-at-Home Orders on Business Closure (LPM)}
\centering
\begin{tabular}[t]{lccc}
\toprule
  & (1) All Data & (2) Ad Subset & (3) No Ad Subset\\
\midrule
Construction & \num{0.009} (\num{0.020}) & \num{-0.027} (\num{0.030}) & \num{0.065} (\num{0.017})***\\
Retail and wholesale & \num{0.003} (\num{0.020}) & \num{-0.031} (\num{0.030}) & \num{0.060} (\num{0.016})***\\
Services & \num{0.042} (\num{0.020})** & \num{0.015} (\num{0.030}) & \num{0.089} (\num{0.016})***\\
Hotels, cafes, and restaurants & \num{0.050} (\num{0.020})** & \num{0.026} (\num{0.030}) & \num{0.090} (\num{0.017})***\\
Information and communication & \num{0.005} (\num{0.020}) & \num{-0.024} (\num{0.030}) & \num{0.056} (\num{0.017})***\\
Manufacturing & \num{0.014} (\num{0.020}) & \num{-0.021} (\num{0.030}) & \num{0.070} (\num{0.017})***\\
Agriculture & \num{0.004} (\num{0.041}) & \num{-0.054} (\num{0.061}) & \num{0.098} (\num{0.034})***\\
Transportation and logistics & \num{0.049} (\num{0.021})** & \num{0.019} (\num{0.032}) & \num{0.095} (\num{0.020})***\\

\bottomrule
\multicolumn{4}{l}{\rule{0pt}{1em}Std. errors are clustered at the business level, * p $<$ 0.1, ** p $<$ 0.05, *** p $<$ 0.01}\\
\end{tabular}
\label{table:staggered_lpm}
\end{table}

\begin{table}

\caption{Effect of Stay-at-Home Orders on Business Closure (Logit)}
\centering
\begin{tabular}[t]{lccc}
\toprule
  & (1) All Data & (2) Ad Subset & (3) No Ad Subset\\
\midrule
Construction & \num{0.013} (\num{0.020}) & \num{-0.017} (\num{0.019}) & \num{0.060} (\num{0.032})*\\
Retail and wholesale & \num{0.008} (\num{0.019}) & \num{-0.020} (\num{0.018}) & \num{0.055} (\num{0.032})*\\
Services & \num{0.046} (\num{0.026})* & \num{0.024} (\num{0.027}) & \num{0.084} (\num{0.036})**\\
Hotels, cafes, and restaurants & \num{0.055} (\num{0.027})** & \num{0.037} (\num{0.029}) & \num{0.086} (\num{0.035})**\\
Information and communication & \num{0.009} (\num{0.019}) & \num{-0.014} (\num{0.019}) & \num{0.051} (\num{0.030})*\\
Manufacturing & \num{0.015} (\num{0.021}) & \num{-0.015} (\num{0.020}) & \num{0.064} (\num{0.033})*\\
Agriculture & \num{0.002} (\num{0.039}) & \num{-0.063} (\num{0.060}) & \num{0.054} (\num{0.036})\\
Transportation and logistics & \num{0.052} (\num{0.028})* & \num{0.027} (\num{0.030}) & \num{0.090} (\num{0.038})**\\
\bottomrule
\multicolumn{4}{l}{\rule{0pt}{1em}Std. errors are clustered at the business level, * p $<$ 0.1, ** p $<$ 0.05, *** p $<$ 0.01}\\
\end{tabular}
\label{table:staggered_logit}
\end{table}

\clearpage

\section*{Web Appendix A: OxCGRT}

The Oxford Covid-19 Government Response Tracker (OxCGRT) contains information on government policies implemented in response to the pandemic. The specific indicators are displayed in Table A1, and the components of the indices are displayed in Table A2.

\setcounter{table}{0}
\renewcommand{\thetable}{A\arabic{table}}

\begin{table}[ht]
\centering
\footnotesize
\caption{Name and Description of OxCGRT Indicators} 
\begin{tabular}
{|p{0.5cm}|p{4cm}|p{7cm}|}
  \hline
ID & Name & Description \\ 
  \hline
C1 & C1\_School closing & Record closings of schools and universities \\ 
  C2 & C2\_Workplace closing & Record closings of workplaces \\ 
  C3 & C3\_Cancel public events & Record cancelling of public events \\ 
  C4 & C4\_Restrictions on gatherings & Record limits on gatherings \\ 
  C5 & C5\_Close public transport & Record closing of public transport \\ 
  C6 & C6\_Stay at home requirements & Record orders to shelter in place and otherwise be confined to the home \\ 
  C7 & C7\_Restrictions on internal movement & Record restrictions on internal movement between cities/regions \\ 
  C8 & C8\_International travel controls & Record restrictions on international travel Note: this records policy for foreign travelers, not citizens \\ 
  \hline
  E1 & E1\_Income support (for households) & Record if the government is providing direct cash payments to people who lose their jobs or cannot work. Note: only includes payments to firms if explicitly linked to payroll/salaries \\ 
  E2 & E2\_Debt/contract relief (for households) & Record if the government is freezing financial obligations for households (e.g., stopping loan repayments, preventing services such as water from stopping, or banning evictions) \\ 
  E3 & E3\_Fiscal measures & Announced economic stimulus spending Note: only record amount additional to previously announced spending \\ 
  E4 & E4\_International support & Announced offers of COVID-19 related aid spending to other countries Note: only record amount additional to previously announced spending \\ 
  \hline
  H1 & H1\_Public information campaigns & Record presence of public information campaigns \\ 
  H2 & H2\_Testing policy & Record government policy on who has access to testing Note: this records policies about testing for current infection (PCR tests) not testing for immunity (antibody test) \\ 
  H3 & H3\_Contact tracing & Record government policy on contact tracing after a positive diagnosis Note: policies that would identify all people potentially exposed to COVID-19; voluntary bluetooth apps are unlikely to achieve this \\ 
  H4 & H4\_Emergency investment in healthcare & Announced short-term spending on healthcare system (e.g., hospitals, masks) Note: only record amount additional to previously announced spending \\ 
  H5 & H5\_Investment in vaccines & Announced public spending on COVID-19 vaccine development Note: only record amount additional to previously announced spending \\ 
  H6 & H6\_Facial Coverings & Record policies on the use of facial coverings outside the home \\ 
  H7 & H7\_Vaccination Policy & Record policies for vaccine delivery for different groups \\ 
   \hline
   \multicolumn{3}{l}{\footnotesize{See Hale et al. (2020) for details on the coding of each indicator.}} \\ 
\end{tabular}
\end{table}

\begin{table}[ht]
\centering
\caption{Components of OxCGRT Indices} 
\begin{tabular}{rccc}
  \hline
  \hline
ID & Government & Containment & Stringency \\ 
 & response & health & index \\ 
 & index & index &  \\ 
 \hline
  C1 & x & x & x \\ 
  C2 & x & x & x \\ 
  C3 & x & x & x \\ 
  C4 & x & x & x \\ 
  C5 & x & x & x \\ 
  C6 & x & x & x \\ 
  C7 & x & x & x \\ 
  C8 & x & x & x \\ 
  E1 & x &  &  \\ 
  E2 & x &  &  \\ 
  E3 &  &  &  \\ 
  E4 &  &  &  \\ 
  H1 & x & x & x \\ 
  H2 & x & x &  \\ 
  H3 & x & x &  \\ 
  H4 &  &  &  \\ 
  H5 &  &  &  \\ 
  H6 & x & x &  \\ 
  H7 & x & x &  \\ 
   \hline
\multicolumn{4}{l}{\footnotesize{See Hale et al. (2020) for details on the coding on each}} \\ 
\multicolumn{4}{l}{\footnotesize{indicator.}} \\ 
\end{tabular}
\end{table}

\clearpage

\section*{Web Appendix B: Results by Country}

\setcounter{table}{0}
\renewcommand{\thetable}{B\arabic{table}}

The following tables present the results of government lockdowns and advertising by country.

\begin{table}[ht]
\centering
\footnotesize
\caption{Estimated Effect of Government Lockdowns by Country} 
\begin{tabular}{lrrr}
  \hline
Country & Estimate & Lower & Upper \\ 
  \hline
India & 0.1588 & 0.0665 & 0.2520 \\ 
  Bangladesh & 0.1445 & 0.0457 & 0.2651 \\ 
  Uganda & 0.1355 & 0.0573 & 0.2122 \\ 
  Egypt & 0.1309 & 0.0330 & 0.2634 \\ 
  Malaysia & 0.1278 & 0.0282 & 0.2767 \\ 
  United Kingdom & 0.1266 & 0.0439 & 0.2259 \\ 
  United Arab Emirates & 0.1241 & 0.0359 & 0.2368 \\ 
  Philippines & 0.1240 & 0.0353 & 0.2490 \\ 
  Nigeria & 0.1232 & 0.0326 & 0.2503 \\ 
  South Africa & 0.1230 & 0.0385 & 0.2327 \\ 
  Pakistan & 0.1201 & 0.0489 & 0.1903 \\ 
  Argentina & 0.1176 & 0.0467 & 0.2028 \\ 
  Colombia & 0.1137 & 0.0478 & 0.1886 \\ 
  Israel & 0.1109 & 0.0342 & 0.2105 \\ 
  Peru & 0.1107 & 0.0537 & 0.1693 \\ 
  Kenya & 0.1094 & 0.0458 & 0.1771 \\ 
  Ghana & 0.1085 & 0.0319 & 0.2058 \\ 
  Australia & 0.1070 & 0.0246 & 0.2361 \\ 
  Saudi Arabia & 0.0970 & 0.0200 & 0.2214 \\ 
  Vietnam & 0.0952 & 0.0134 & 0.2575 \\ 
  United States & 0.0921 & 0.0309 & 0.1679 \\ 
  Mexico & 0.0911 & 0.0347 & 0.1533 \\ 
  France & 0.0842 & 0.0237 & 0.1688 \\ 
  Canada & 0.0831 & 0.0289 & 0.1489 \\ 
  Ecuador & 0.0828 & 0.0374 & 0.1325 \\ 
  Romania & 0.0815 & 0.0250 & 0.1541 \\ 
  Turkey & 0.0815 & 0.0280 & 0.1422 \\ 
  Spain & 0.0793 & 0.0295 & 0.1390 \\ 
  Brazil & 0.0774 & 0.0202 & 0.1621 \\ 
  Greece & 0.0721 & 0.0165 & 0.1607 \\ 
  Germany & 0.0711 & 0.0244 & 0.1281 \\ 
  Thailand & 0.0669 & 0.0156 & 0.1468 \\ 
  Italy & 0.0643 & 0.0173 & 0.1327 \\ 
  Cambodia & 0.0617 & 0.0149 & 0.1316 \\ 
  Portugal & 0.0611 & 0.0256 & 0.1002 \\ 
  Indonesia & 0.0548 & 0.0166 & 0.1063 \\ 
  Belgium & 0.0542 & 0.0142 & 0.1162 \\ 
  Netherlands & 0.0508 & 0.0163 & 0.0940 \\ 
  Denmark & 0.0481 & 0.0084 & 0.1232 \\ 
  Czechia & 0.0463 & 0.0095 & 0.1119 \\ 
  Switzerland & 0.0431 & 0.0117 & 0.0885 \\ 
  Norway & 0.0379 & 0.0090 & 0.0845 \\ 
  Hungary & 0.0351 & 0.0133 & 0.0594 \\ 
  South Korea & 0.0326 & 0.0083 & 0.0704 \\ 
  Poland & 0.0315 & 0.0108 & 0.0553 \\ 
  Sweden & 0.0292 & 0.0077 & 0.0628 \\ 
  Japan & 0.0229 & 0.0038 & 0.0616 \\ 
  Singapore & 0.0000 & 0.0000 & 0.0000 \\ 
  Ireland & 0.0000 & 0.0000 & 0.0000 \\ 
   \hline
   \multicolumn{4}{l}{Estimates are calculated from the results of model 3} \\ 
   \multicolumn{4}{l}{in Table 2.} \\ 
\end{tabular}
\label{table:lockdown_by_country}
\end{table}

\begin{table}[ht]
\centering
\footnotesize
\caption{Estimated Effect of Advertising by Country} 
\begin{tabular}{lrrr}
  \hline
Country & Estimate & Lower & Upper \\ 
  \hline
India & -0.0627 & -0.1511 & -0.0158 \\ 
  Israel & -0.0504 & -0.1913 & -0.0024 \\ 
  Bangladesh & -0.0477 & -0.1324 & -0.0075 \\ 
  Malaysia & -0.0471 & -0.1618 & -0.0037 \\ 
  United Kingdom & -0.0420 & -0.1294 & -0.0045 \\ 
  Philippines & -0.0420 & -0.1100 & -0.0095 \\ 
  Argentina & -0.0404 & -0.1115 & -0.0079 \\ 
  United Arab Emirates & -0.0404 & -0.1263 & -0.0048 \\ 
  Pakistan & -0.0401 & -0.1108 & -0.0076 \\ 
  Australia & -0.0397 & -0.1472 & -0.0020 \\ 
  France & -0.0397 & -0.1732 & -0.0019 \\ 
  Vietnam & -0.0394 & -0.1748 & -0.0014 \\ 
  Italy & -0.0387 & -0.2060 & -0.0009 \\ 
  Peru & -0.0384 & -0.0999 & -0.0090 \\ 
  Egypt & -0.0379 & -0.1163 & -0.0040 \\ 
  South Africa & -0.0369 & -0.1081 & -0.0054 \\ 
  United States & -0.0366 & -0.1415 & -0.0023 \\ 
  Uganda & -0.0332 & -0.0644 & -0.0132 \\ 
  Colombia & -0.0330 & -0.0773 & -0.0094 \\ 
  Spain & -0.0319 & -0.1246 & -0.0023 \\ 
  Saudi Arabia & -0.0315 & -0.1075 & -0.0038 \\ 
  Canada & -0.0313 & -0.1203 & -0.0023 \\ 
  Kenya & -0.0309 & -0.0746 & -0.0082 \\ 
  Germany & -0.0302 & -0.1303 & -0.0015 \\ 
  Romania & -0.0291 & -0.1119 & -0.0022 \\ 
  Nigeria & -0.0290 & -0.0715 & -0.0071 \\ 
  Greece & -0.0282 & -0.1261 & -0.0012 \\ 
  Ecuador & -0.0274 & -0.0815 & -0.0049 \\ 
  Turkey & -0.0269 & -0.0949 & -0.0027 \\ 
  Ghana & -0.0257 & -0.0680 & -0.0056 \\ 
  Denmark & -0.0256 & -0.1463 & -0.0004 \\ 
  Brazil & -0.0243 & -0.0913 & -0.0021 \\ 
  Belgium & -0.0229 & -0.1043 & -0.0012 \\ 
  Portugal & -0.0228 & -0.0854 & -0.0023 \\ 
  Czechia & -0.0200 & -0.0954 & -0.0009 \\ 
  Thailand & -0.0198 & -0.0806 & -0.0013 \\ 
  Netherlands & -0.0190 & -0.0807 & -0.0013 \\ 
  Switzerland & -0.0188 & -0.0908 & -0.0008 \\ 
  Norway & -0.0170 & -0.0895 & -0.0005 \\ 
  Cambodia & -0.0165 & -0.0688 & -0.0010 \\ 
  Mexico & -0.0157 & -0.0443 & -0.0029 \\ 
  Hungary & -0.0146 & -0.0664 & -0.0009 \\ 
  Indonesia & -0.0126 & -0.0330 & -0.0031 \\ 
  South Korea & -0.0125 & -0.0524 & -0.0010 \\ 
  Sweden & -0.0125 & -0.0694 & -0.0003 \\ 
  Poland & -0.0104 & -0.0474 & -0.0006 \\ 
  Japan & -0.0087 & -0.0448 & -0.0004 \\ 
  Singapore & -0.0000 & -0.0000 & -0.0000 \\ 
  Ireland & -0.0000 & -0.0000 & -0.0000 \\ 
   \hline
   \multicolumn{4}{l}{Estimates are calculated from the results of model 3} \\ 
   \multicolumn{4}{l}{in Table 2.} \\ 
\end{tabular}
\label{table:ad_by_country}
\end{table}

\end{document}